\documentclass[twocolumn,showpacs,superscriptaddress,twoside,prl]{revtex4-1}
\usepackage{amsthm}
\usepackage{amsmath}
\usepackage{amssymb}
\usepackage{graphicx}
\usepackage{url}
\usepackage{dsfont}
\usepackage{float}
\usepackage{bm}
\usepackage{ifthen}
\usepackage[usenames,dvipsnames]{color}
\usepackage{mathrsfs}

\usepackage[colorlinks=true,citecolor=blue,urlcolor=black]{hyperref}
\usepackage{float}
\usepackage{afterpage}
\usepackage{subfigure}
\usepackage{adjustbox}
\usepackage{times}

\newcommand{\sket}[1]{{\ensuremath{\lvert#1\rangle}}}
\newcommand{\lket}[1]{{\ensuremath{\left\lvert#1\right\rangle}}}
\newcommand{\ket}[1]{\if@display\lket{#1}\else\sket{#1}\fi}
\newcommand{\sbra}[1]{{\ensuremath{\langle#1\rvert}}}
\newcommand{\lbra}[1]{{\ensuremath{\left\langle#1\right\rvert}}}
\newcommand{\bra}[1]{\if@display\lbra{#1}\else\sbra{#1}\fi}
\newcommand{\sbraket}[2]{{\ensuremath{\langle#1\rvert#2\rangle}}}
\newcommand{\lbraket}[2]{{\ensuremath{\left\langle#1\!\left\rvert\vphantom{#1}#2\right.\!\right\rangle}}}
\newcommand{\braket}[2]{\if@display\lbraket{#1}{#2}\else\sbraket{#1}{#2}\fi}

\newcommand{\sketbra}[2]{{\ensuremath{\lvert #1\rangle\!\langle #2\rvert}}}
\newcommand{\lketbra}[2]{{\ensuremath{\left\lvert #1\right\rangle\!\!\left\langle #2\right\rvert}}}
\newcommand{\ketbra}[2]{\if@display\lketbra{#1}{#2}\else\sketbra{#1}{#2}\fi}

\theoremstyle{plain}

\theoremstyle{definition}

\usepackage{color}
\definecolor{myred}{rgb}{1,0,0}

\definecolor{myblue}{rgb}{0,0,0.8}

\definecolor{myyellow}{rgb}{0.9,0.8,0}

\definecolor{mygreen}{rgb}{0,0.6,0}

\definecolor{myorange}{rgb}{0.6,0.6,0}

\definecolor{mycerul}{rgb}{0,0.6,1}

\newcommand{\be}{\begin{equation}}
\newcommand{\ee}{\end{equation}}

\newcommand{\ignore}[1]{} 

\begin{document}

\title{  Qudit-Basis  Universal Quantum Computation using  $ \chi^{(2)} $ Interactions  }
\author{Murphy Yuezhen Niu}
\affiliation{Research Laboratory of Electronics, Massachusetts Institute of Technology, Cambridge, Massachusetts 02139, USA}
\affiliation{Department of Physics, Massachusetts Institute of Technology, Cambridge, Massachusetts 02139, USA}
\author{Isaac L. Chuang}
\affiliation{Research Laboratory of Electronics, Massachusetts Institute of Technology, Cambridge, Massachusetts 02139, USA}
\affiliation{Department of Physics, Massachusetts Institute of Technology, Cambridge, Massachusetts 02139, USA}
\affiliation{Department of Electrical Engineering and Computer Science, Massachusetts Institute of Technology, Cambridge, Massachusetts 02139, USA}
\author{Jeffrey H. Shapiro}
\affiliation{Research Laboratory of Electronics, Massachusetts Institute of Technology,  Cambridge, Massachusetts 02139, USA}
\affiliation{Department of Electrical Engineering and Computer Science, Massachusetts Institute of Technology, Cambridge, Massachusetts 02139, USA}

\date{\today}
\begin{abstract}
We prove that universal quantum computation can be realized---using only linear optics and $\chi^{(2)}$ (three-wave mixing) interactions---in any $(n+1)$-dimensional qudit basis of the $n$-pump-photon subspace.  First, we exhibit a strictly universal gate set for the qubit basis in the one-pump-photon subspace.  Next, we demonstrate qutrit-basis universality by proving that $\chi^{(2)}$ Hamiltonians and photon-number operators generate the full $\mathfrak{u}(3)$ Lie algebra in the two-pump-photon subspace, and showing how the qutrit controlled-$Z$ gate can be implemented with only linear optics and $\chi^{(2)}$ interactions. We then use proof by induction to obtain our general qudit result.  Our induction proof relies on coherent photon injection/subtraction, a technique enabled by $\chi^{(2)}$ interaction between the encoding modes and ancillary modes.  Finally, we show that coherent photon injection is more than a conceptual tool in that it offers a route to preparing high-photon-number Fock states from single-photon Fock states.
\end{abstract} 
\maketitle

\textit{Introduction.}--- Photons are promising information carriers for quantum computers owing to photons' long room-temperature coherence time,  high transmission speed,  and high-fidelity preparation schemes~\cite{yuan2002electrically,moreau2001single,babinec2010diamond, Wong2017}, plus the availability of efficient photodetectors~\cite{gol2001picosecond, pernice2012high}, and the scalable on-chip integration of linear and nonlinear optical components~\cite{o2007optical,politi2008silica, claudon2010highly,sprengers2011waveguide}.  Architectures for optics-based quantum computation have gone through dramatic developments over the past two decades~\cite{Weinfurter1995,Chuang1995, KLM, kok2007linear, Girvin2016new,Lloyd1999}, but significant obstacles remain to be overcome.

Optics-based quantum computation depends on photon-photon interactions for the realization of a universal gate set. The lowest order photon-photon interactions are described by unitary transformations  of the form $\hat{U}=\exp(-i\hat{L})$ that are generated by general two-wave mixing Hamiltonians,
\begin{equation}
\hat{L}  \in \{ (g \hat{a}\hat{b}+g^* \hat{a}^\dagger\hat{b}^\dagger),  (g \hat{a}\hat{b}^\dagger+g^* \hat{a}^\dagger\hat{b}) \},
\label{twowavemix}
\end{equation}
where  $g$ is a $c$-number and $\hat{a}^\dagger$ and $\hat{b}^\dagger$ are photon-creation operators from different optical modes, so that $[\hat{a},\hat{b}^\dagger] = 0$, or the same optical mode, for which $[\hat{a},\hat{b}^\dagger] = 1$. Unitary transformations of this form can realize universal single-qubit rotations in the Fock-state basis but are  not universal for  quantum computation without some additional resource.   To implement universal optics-based quantum computation, four-wave mixing~(a $\chi^{(3)}$ interaction) was previously considered to be the lowest-order optical nonlinearity that will suffice in this regard~\cite{Milburn1989,Lloyd1999,Jeff2012}. The inherent weakness of  $\chi^{(3)}$ interactions, however, has precluded their delivering the  high-fidelity gates required to make optics-based quantum computation practical~\cite{shapiro2006single,shapiro2007,chudzicki2013deterministic}. Linear-optical quantum computation (LOQC)~\cite{KLM,raussendorf2001one,barrett2005efficient,browne2005resource} circumvents the need for photon-photon interactions through postselection, but this approach comes with the need for a prohibitive number of perfect single-photon ancillae to cope with  LOQC's probabilistic nature and the ubiquitous  photon loss~\cite{kok2007linear,Zaidi2015,Rudolph2015,Pant2016}. 

One way to circumvent the weakness of photon-photon interactions  is to employ  the \textit{lowest}-order nonlinearity that can provide universal quantum computation, viz.,  the $\chi^{(2)}$ interaction whose three-wave-mixing Hamiltonians can be decomposed into linear combinations of the following terms~\cite{chi2type}
\begin{align}\label{HSPDC11}
	\hat{G}_1
		&=\frac{i\kappa}{2}\left[\hat{a}_s^{\dagger}\hat{a}_i^{\dagger}\hat{a}_p - \hat{a}_s\hat{a}_i\hat{a}_p^{\dagger}\right],
		\,\hat{G}_2
		&=\frac{\kappa}{2}\left[\hat{a}_s^{\dagger}\hat{a}_i^{\dagger}\hat{a}_p + \hat{a}_s\hat{a}_i\hat{a}_p^{\dagger}\right].
\end{align}
Here, $\{\hat{a}_k^{\dagger}: k = s,i,p\}$
are the photon-creation operators of the interaction's signal, idler, and pump modes, and the real-valued $\kappa$ quantifies the interaction's strength. 

The efficiencies of  $\chi^{(2)}$ interactions have been steadily improving  over the past decade~\cite{Pan2010experimental,barz2010,Ramelow2011,guerreiro2013interaction,hamel2014direct,Jennewein2014,meyer2015power,Niu2016UPDC,Long2008,Devoret2010,Bergeal2010,Tian2011,Tian2012,Oliver2016}. Moreover, owing to the importance of $\chi^{(2)}$ interactions in quantum state transduction for superconducting and ion-trap qubits, the platforms of interest for $\chi^{(2)}$ interactions have expanded beyond traditional nonlinear crystals~\cite{Long2008,Devoret2010,Bergeal2010,Tian2011,Tian2012,Oliver2016}, bringing full utilization of their quantum dynamics closer to reality.  

Coherent photon conversion, i.e., $\chi^{(2)}$ interactions defined in (\ref{HSPDC11}) in which the signal, idler, and pump modes are all quantum mechanical, was first proposed by Koshino~\cite{Koshino2009}, and later used by Langford \emph{et al}.~\cite{Ramelow2011} to show how universal quantum computation can be realized with that resource in the single-photon qubit basis. We refer to such interactions as \emph{full-quantum} $\chi^{(2)}$ interactions, to distinguish them from \emph{pumped}  $\chi^{(2)}$ interactions, in which a nondepleting coherent-state pump reduces (\ref{HSPDC11})  to the two-wave interactions shown in (\ref{twowavemix}).  Langford \emph{et al}.'s groundbreaking work, however, is not without drawback.  Available  schemes for correcting photon loss~\cite{Chuang1995,Ralph2004,Loock2016,Girvin2016}, viz., the dominant error in photonic quantum computation, require either measurement-based or $\chi^{(3)}$ gates on the encoded basis.  Thus Ref. \cite{Ramelow2011} does  not provide a  $\chi^{(2)}$ approach that facilitates photonic quantum computation that is robust to photon loss.     
 
In this Letter, and its companion paper~\cite{ToBepublished}, we show how the work of Langford \emph{et al}.\@ can be extended to a more natural computational basis for $\chi^{(2)}$-based quantum computation in which photon-loss errors can be addressed.  More generally, we prove that $\chi^{(2)}$ interactions plus linear optics can provide a strictly universal gate set for quantum computation in any $(n+1)$-dimensional qudit basis of the $n$-pump-photon subspace. Because any  $d$-qudit unitary gate can be described by a Lie group element of SU$((n+1)^d)$, the universality of a given class of Hamiltonians is directly related to that class's Lie algebra and the Lie group it generates via the exponential map~\cite{LieReference}.  Thus we use Lie-algebra analysis to identify code subspaces that are closed under  $\chi^{(2)}$ Hamiltonian evolutions~\cite{Bacon2000,Kemp2001}.  Our  Lie-algebra analysis underlies the symmetry-operator formulation of qudit-basis error-correcting codes for photon-loss errors and the universal gate-set constructions in the encoded basis that we report in~\cite{ToBepublished}. 
Hence our proposal provides a $\chi^{(2)}$ approach to photonic quantum computation that is robust to photon loss.  We begin the development of our universality results with a summary of the linear optics and the  $\chi^{(2)}$ resources we shall employ.  We follow with qubit and qutrit universality proofs, as preludes to our induction proof for the general qudit case.\\ 

\textit{Linear Optics and $\chi^{(2)}$ Resources.}--- The linear optics resources we require are readily available:  dichroic mirrors and phase shifters.  The pumped $\chi^{(2)}$ resource we require is quantum-state frequency conversion (QFC)~\cite{Kumar1990,Albota2004,Albota2006}, which converts a frequency-$\omega_{\rm in}$ single-photon Fock state to a frequency-$\omega_{\rm out}$ single-photon Fock state.  The full-quantum $\chi^{(2)}$ resources we require are: second-harmonic generation (SHG), which converts a frequency-$\omega_{\rm in}$ two-photon Fock state to a frequency-$2\omega_{\rm in}$ single-photon Fock state; type-I phase-matched spontaneous parametric downconversion (SPDC), which converts a frequency-$2\omega_{\rm in}$ single-photon Fock state to a frequency-$\omega_{\rm in}$ two-photon Fock state; and generalized sum-frequency generation (SFG$_\theta$), which accomplishes the state transformation~\cite{supp}
\begin{align}\label{SFG}
\text{SFG}_\theta \ket{1,1,0} = \cos(\theta)\ket{1,1,0} + \sin(\theta) \ket{0,0,1},
\end{align}
where $\ket{n_s,n_i,n_p}$ denotes a three-mode Fock state containing $n_s$ frequency-$\omega_s$ photons, $n_i$ frequency-$\omega_i$ idler photons, and $n_p$ frequency-$\omega_p$ pump photons, with the pump's frequency satisfying $\omega_p = \omega_s + \omega_i$.\\    

\textit{Universality in the Qubit Basis.}--- 
The Lie group generated by $\chi^{(2)}$ Hamiltonian evolutions is a subgroup of the unitary group $U$, hence it is compact. A compact Lie group, together with its generating Lie algebra, are completely reducible.  This means that they can be written as a direct sum of irreducible representations over the state space $\mathcal{H} \equiv \oplus_{n=1}^\infty \mathcal{H}_n$, whose irreducible subspaces, $\{\mathcal{H}_n\}$, are labeled by their pump mode's maximum photon number $n$, i.e., they are  the $n$-pump-photon subspaces spanned by the three-mode Fock-state bases $\{\ket{0,0,n},\ket{1,1,n-1},\ldots ,  \ket{n,n,0} \}$.  For qubit universality, we therefore encode in the one-pump-photon subspace $\mathcal{H}_1$, using the three-mode Fock states, 
 \begin{align}\label{qubitBasisEq}
&\ket{\tilde{0}}=\ket{1,1,0},\, \,\ket{\tilde{1}}=\ket{0,0,1},
\end{align}
for our logical-qubit basis states.  Here, the signal and idler are both at frequency $\omega$ with orthogonal polarizations, the pump is at frequency $2\omega$, and all three share a common spatial mode.  Universality is proved by the following theorem.  
 
\textit{Theorem 1}. Universal quantum computation can be realized with $\chi^{(2)}$ interactions and linear optics in any qubit basis of the one-pump photon subspace.

\textit{Proof}:  The $\chi^{(2)}$ Hamiltonians, $\hat{G}_1$ and $\hat{G}_2$, defined in (\ref{HSPDC11}) are proportional to the Pauli $\hat{Y}$ and Pauli $\hat{X}$ operators in the logical-qubit basis, which are universal for realizing single-qubit rotations.  So, to complete our $\chi^{(2)}$ universality proof for the logical-qubit basis in (\ref{qubitBasisEq}), it suffices for us to show that we can construct a controlled-$Z$ qubit gate for that basis~\cite{Weinfurter1995}, i.e., a gate (denoted $\Lambda_2[Z]$ in what follows) that imparts a $\pi$-rad phase shift to the $\ket{\tilde{1}}_c\ket{\tilde{1}}_t$ component of the joint state of the control (subscript $c$) and target (subscript $t$) qubits.  Moreover, because $\Lambda_2[Z]$ can be sandwiched between single-qubit $\chi^{(2)}$ rotations to achieve the controlled-$Z$ function in any $\mathcal{H}_1$ qubit basis, Theorem~1 will be proved once we have established how to realize $\Lambda_2[Z]$.

\begin{figure}[H]
\begin{center}
\includegraphics[width=0.85\linewidth]{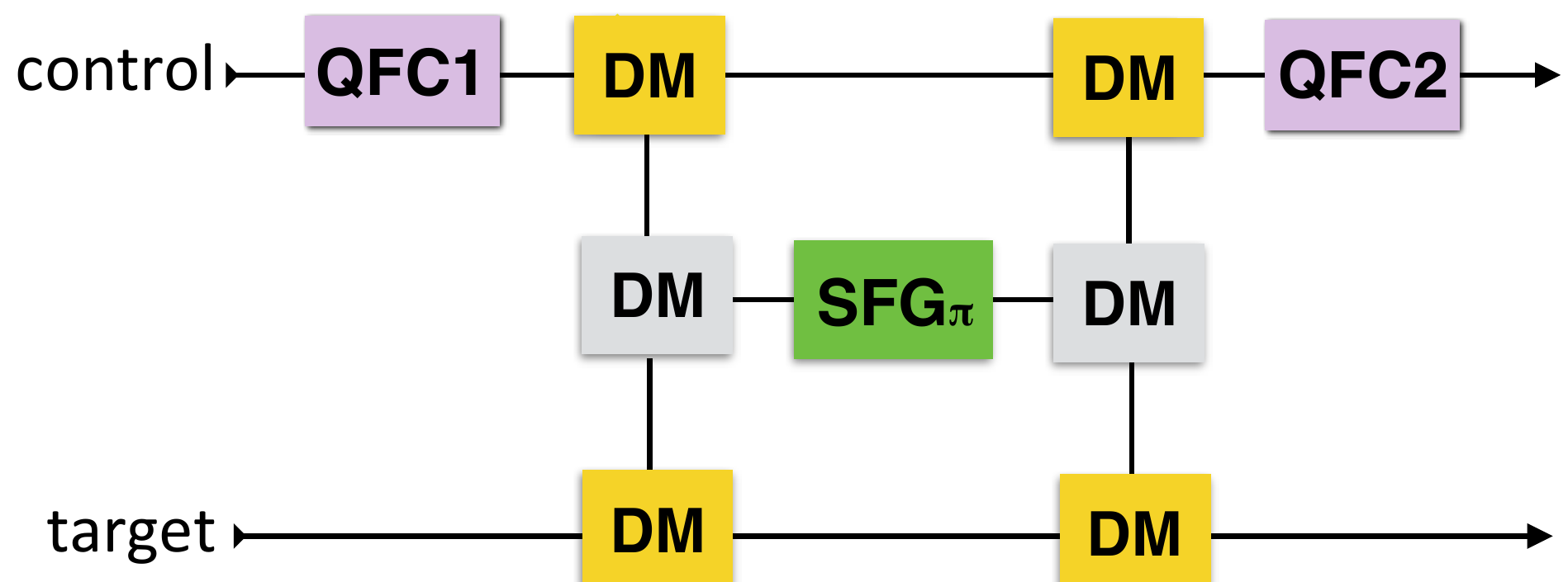}
\caption{Schematic for constructing the $\Lambda_2[Z]$ gate in the logical-qubit basis (\ref{qubitBasisEq}) using $\chi^{(2)}$ interactions and linear optics.  QFC1 and QFC2:  quantum-state frequency conversions.  DM:  dichroic mirror.  SFG$_\pi$:  generalized sum-frequency generation (\ref{SFG}) with $\theta = \pi$.} 
\label{CSGATEGRAPH1}
\end{center}
\end{figure}

Figure~\ref{CSGATEGRAPH1} shows our optical circuit~\cite{footnote} for the $\Lambda_2[Z]$ gate for the logical-qubit basis in (\ref{qubitBasisEq}).  The control and target qubits enter on the upper and lower rails, respectively.  QFC1 shifts the frequency of control qubit's pump photon (if present) from $2\omega$ to $2\omega'$, so that dichroic mirrors (DMs) are able to direct pump photons from the control and target qubits to the center rail's SFG$_\pi$ gate, where they serve as modes~1 (frequency $\omega_1 \equiv 2\omega'$) and 2 (frequency $\omega_2 = 2\omega$).  This gate imparts a $\pi$-rad phase shift if and only if pump photons are present from both the control and target qubits.  Thus, after another set of DMs restore the control and target pump photons to the top and bottom rails, respectively, the $\Lambda_2[Z]$ gate---and hence the proof of Theorem~1---is completed by QFC2, which shifts the frequency of the control qubit's pump photon (if present) from $2\omega'$ to $2\omega$.  Note that each $\chi^{(2)}$ element  Fig.~\ref{CSGATEGRAPH1} acts on only one of its potentially excited bosonic-mode inputs, e.g., QFC1 affects its pump-mode input but neither its signal-mode input nor its idler-mode input. Such modal selectivity puts a burden on experimental realization.  In particular, QFC1 and QFC2 will require a different nonlinear medium than will SFG$_{\rm \pi}$.  This difficulty, however, may disappear once high-efficiency nondepleted $\chi^{(3)}$ induced $\chi^{(2)}$  interactions become available~\cite{Ramelow2011,Jennewein2014,meyer2015power}.

\textit{Universality in the Qutrit Basis.}---  For qutrit universality, we encode in $\mathcal{H}_2$ using the three-mode Fock states
\begin{align}\label{Qutritbasis1}
\ket{\tilde{0}}= \ket{1,1,1}, \,\, \ket{\tilde{1}}=\ket{2,2,0},\,\, \ket{\tilde{2}}=\ket{0,0,2},
\end{align}
for our logical-qutrit basis states.  Here, the signal and idler have frequency $\omega$ and are orthogonally polarized, while the pump has frequency $2\omega$, and all three share a common spatial mode.  These states can be  prepared by type-II phase-matched SPDC in the two-pump-photon subspace~\cite{Niu2016UPDC}, and are naturally confined to this subspace under  $\chi^{(2)}$ interactions. It follows that restricting linear combinations of the $\chi^{(2)}$ Hamiltonians, $\hat{G}_1, \hat{G}_2$, the modal photon-number operators, $\{\hat{N}_k \equiv \hat{a}_k^\dagger\hat{a}_k : k = s, i, p\}$, and the nested commutators of these operators to the two-pump-photon subspace $\mathcal{H}_2$ constitutes a Lie algebra $\mathfrak{g}$.  The Lie group $H$ associated with $\mathfrak{g}$ is found from the exponential map $\exp: \mathfrak{g}\to H$, where  for each group element $\hat{h}\in H$, $\exists  \hat{E}\in  \mathfrak{g}$  and $t \in  \mathbb{R}$ such that  $\hat{h}=\exp( it \hat{E})$. For simplicity, in all that follows, we set $\kappa = 1$ in the Hamiltonians $\hat{G}_1$ and $\hat{G}_2$.  We begin our universality demonstration with a theorem about $\mathfrak{g}$.

\textit{Theorem 2}. The Lie algebra $\mathfrak{g}$ is $ \mathfrak{u}(3) $.

\textit{Proof}: First we prove that $ \mathfrak{u}(3) \subseteq \mathfrak{g} $. From the original $\chi^{(2)}$ Hamiltonians $\hat{G}_1$ and $\hat{G}_2$, we can obtain all transformations generated by linear combinations of $\hat{G}_1$, $\hat{G}_2$, $\hat{N}_s$, $\hat{N}_i$, $\hat{N}_p$ and their nested commutators.  
Using the vector  $ \bf{v}^T\equiv [\begin{array}{ccc} v_0 &  v_1 & v_2 \end{array}]$ to represent the qutrit $\ket{\psi}=v_0\ket{1,1,1}+ v_1 \ket{2,2,0} +v_2\ket{0,0,2}$, we obtain the matrix representations
\begin{align}\label{G1}
&\hat{G}_1  =\frac{i}{2}\left[\hat{a}_s^{\dagger}\hat{a}_i^{\dagger}\hat{a}_p - \hat{a}_s\hat{a}_i\hat{a}_p^{\dagger}\right]=\frac{-i}{2}\begin{bmatrix}
		0 &2&\sqrt{2}\\
	-2 & 0&0\\
	-\sqrt{2}& 0&0
\end{bmatrix}, \\
& \hat{G}_2=\frac{1}{2}\left[\hat{a}_s^{\dagger}\hat{a}_i^{\dagger}\hat{a}_p + \hat{a}_s\hat{a}_i\hat{a}_p^{\dagger}\right]=\frac{1}{2}\begin{bmatrix}
		0 &2&\sqrt{2}\\
	2 & 0&0\\
	\sqrt{2}& 0&0
\end{bmatrix},\\\label{G3}
&\hat{G}_3=i[\hat{G}_1, \hat{G}_2]=\begin{bmatrix}
		1 &0&0\\
	0  & -2&0\\
	0& 0&1
\end{bmatrix},\\
&\hat{G}_4= i[\hat{G}_2,\hat{G}_3]=3\begin{bmatrix}
		0 &1&0\\
	1  & 0&0\\
	0&0&0
\end{bmatrix}, \\
& \hat{G}_5= i[\hat{G}_3,\hat{G}_1]=3i\begin{bmatrix}
		0 &1&0\\
	-1  & 0&0\\
	0&0&0
\end{bmatrix},\\\label{G7}
&\hat{G}_6=\frac{1}{2}\left( i[\hat{G}_1, \hat{G}_4] +i[\hat{G}_5, \hat{G}_2] \right)=\frac{3}{4}\begin{bmatrix}
		0 &0&0\\
	0  & 0&1\\
	0&1&0
\end{bmatrix},\\
& \hat{G}_7=i[\hat{G}_4, \hat{G}_2]=\frac{3i}{4}\begin{bmatrix}
		0 &0&0\\
	0  & 0&-1\\
	0&1&0
\end{bmatrix},\\
&\hat{G}_8=\frac{1}{2}(1-\hat{N}_p) =\frac{1}{2}\begin{bmatrix}
		0 &0&0\\
	0  & 1&0\\
	0&0&-1
\end{bmatrix},
\end{align}
\begin{align}
& \hat{G}_9= \frac{1}{2}\!\left(\frac{\hat{N}_s+\hat{N}_i}{2} +\hat{N}_p\right) =\begin{bmatrix}
		1 &0&0\\
	0  & 1&0\\
	0&0&1
\end{bmatrix},
\end{align}
for all the independent generators, where the second equalities apply in the two-pump-photon subspace $\mathcal{H}_2$. It is then straightforward to verify that the Gell-Mann matrices arising from linear combinations of the above generators are:
\begin{align}
&\hat{\lambda}_1=\hat{G}_4/3,  \, &\hat{\lambda}_2=-\hat{G}_5/3, \\
&\hat{\lambda}_3= 2\hat{G}_8 + \hat{G}_3, \, & \hat{\lambda}_4=\sqrt{2}(\hat{G}_2-\hat{G}_4/3),\\
&\hat{\lambda_5}=\sqrt{2}(\hat{G}_1-\hat{G}_5/3), \, & \hat{\lambda}_6= 4\hat{G}_6/3, \\
 & \hat{\lambda}_7= 4\hat{G}_7/3,\, & \hat{\lambda}_8= (\hat{G}_3 +6 \hat{G}_8)/\sqrt{3}.
\end{align}
Gell-Mann matrices are one representation of the complete set of  linearly independent generators for  the $\mathfrak{su}(3)$ Lie algebra.  Together with $\hat{G}_9$ they form the complete set of generators for $\mathfrak{u}(3)$, proving that $\mathfrak{u}(3) \subseteq \mathfrak{g}.$

We complete our proof of Theorem~2 by showing that  $\mathfrak{g} \subseteq \mathfrak{u}(3)$.  Because the two-pump-photon subspace $\mathcal{H}_2$ is closed under  $\mathfrak{g}$,  every Lie group element $\hat{h}\in H$ is generated by an $\hat{E}\in \mathfrak{g}$ via $\hat{h}=\exp(it\hat{E})$ for some $t\in \mathbb{R}$. As $\exp(it\hat{E})$ is a unitary transformation in the two-pump-photon subspace, we have $H\subset U(3)$.  Furthermore, this condition holds if and only if $\mathfrak{g} \subseteq \mathfrak{u}(3)$, thus finishing Theorem~2's proof.

Refs.~\cite{Divincenzo1995,Deutsch1995,Lloyd1996universal} show that if operators $\{\hat{G}_k \}$ and their nested commutators generate the Lie algebra $\mathfrak{u}(3^m)$, then they can be used to construct a universal set of unitaries $U_k(t)=\exp(-it\hat{G}_k)$ in the $m$-qutrit subspace. Setting $m=1$ we have the following claim.

\textit{Claim 1}. Universal single-qutrit rotations can be realized with  $\chi^{(2)}$ interactions. 

Universal qutrit computation entails not only universal single-qutrit unitary gates but also universal two-qutrit unitary transformations in $\mathcal{H}_2^{\otimes 2}$, so we need the following theorem.

\textit{Theorem 3}. Universal qutrit quantum computation can be realized with $\chi^{(2)}$ interactions and linear optics in any qutrit basis of the two-pump-photon subspace. 

\textit{Proof}: From Claim~1 we know that arbitrary $U(3)$  qutrit rotations can be realized with  $\chi^{(2)}$ interactions. It is also known~\cite{Weinfurter1995,Muthukrishnan2000multivalued,Vlasov2002noncommutative,Brylinski2002universal} that a universal single-qutrit gate set plus a controlled-$Z$ gate for the logical-qutrit basis in (\ref{Qutritbasis1})---denoted $\Lambda_3[Z]$---are universal for qutrit computation in any qutrit basis of the two-pump-photon subspace $\mathcal{H}_2$.  

\begin{figure}
\begin{center}
\includegraphics[width=\linewidth]{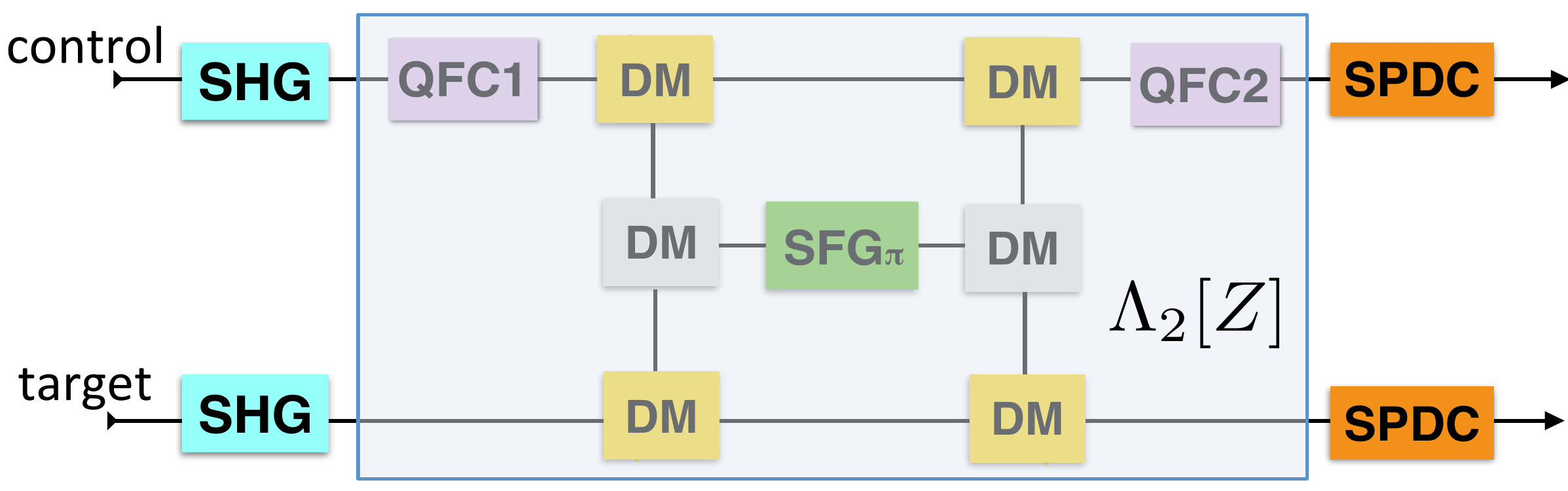} 
\caption{Schematic for constructing the $\Lambda_3[Z]$ gate in the logical-qutrit basis (\ref{Qutritbasis1}) using $\chi^{(2)}$ interactions and linear optics.  SHG:  second-harmonic generation; $\Lambda_2[Z]$:  the optical circuit from Fig.~(\ref{CSGATEGRAPH1}) with modifications described in the text. SPDC:  type-I phase-matched spontaneous parametric downconversion. \label{GateCircuit}}
\end{center} 
\end{figure}

The $\Lambda_3[Z]$ gate realizes the unitary transformation $\Lambda_3[Z] \ket{\tilde{j}}_c\ket{\tilde{k}}_t=(-1)^{ \delta_{\tilde{j}\tilde{2}}\delta_{\tilde{k}\tilde{2}}}\ket{\tilde{j}}_c\ket{\tilde{k}}_t$ for states in $\mathcal{H}_2$, where $\delta_{uv}$ is the Kronecker delta.  Figure~\ref{GateCircuit} shows how this gate can be realized using $\chi^{(2)}$ interactions and linear optics.  The control and target qubits enter on the upper and lower rails, respectively, where second-harmonic generators (SHGs) convert two-photon Fock-state pumps at frequency $2\omega$ to a single-photon Fock state at frequency $4\omega$.  The shaded block labeled $\Lambda_2[Z]$ is the same gate shown in Fig.~\ref{CSGATEGRAPH1} \emph{except} that:  (1) its QFC1 converts a frequency-$4\omega$ single-photon Fock state to a frequency $4\omega'$ single-photon Fock state; (2) its first set of DMs route the frequency-$4\omega'$ photon (if present) from the upper rail and the frequency-$4\omega$ photon (if present) from the lower rail to the SFG$_\pi$ block on the center rail; (3) its SFG$_\pi$ block is arranged to apply a $\pi$-rad phase shift to the state $\ket{1,1,0}$, whose first two entries are the photon numbers of its frequency-$4\omega'$ and frequency-$4\omega$ inputs; (4) its second set of DMs return the frequency-$4\omega'$ and frequency-$4\omega$ photons to the upper and lower rails, respectively; and (5) its QFC2 converts a frequency-$4\omega'$ single-photon Fock state to a frequency-$4\omega$ single-photon Fock state.   The SPDC blocks then complete the $\Lambda_3[Z]$ gate---by converting frequency-$4\omega$ single-photon Fock states (if present) to frequency-$2\omega$ two-photon Fock states---because the $\Lambda_2[Z]$ block has imparted a $\pi$-rad phase shift to the $\ket{\tilde{2}}_c\ket{\tilde{2}}_t$ component of the original input state.  Together with Claim~1, the $\Lambda_3[Z]$ construction proves Theorem~3.  \\

\textit{Universality in the $(n+1)$-Dimensional Qudit Basis.}--- The culmination of our $\chi^{(2)}$ universality work is the following theorem.

\textit{Theorem 4.} Universal qudit quantum computation can be realized with $\chi^{(2)}$ interactions and linear optics in any $(n+1)$-dimensional basis of the $n$-pump-photon subspace.

\textit{Proof}:  Our proof is by induction.  We have already shown that Theorem~4 holds for $n=1$ and $n=2$.  The induction proof is completed by assuming that Theorem~4 holds for $n=m$, and then showing that it holds for $n = m+1$.  The details appear in~\cite{supp}.  Here we just note that they involve a Lie-group result~\cite{Kemp2001,Freedman2002} and coherent photon injection/subtraction.  Coherent photon injection/subtraction are full-quantum $\chi^{(2)}$ interactions between the encoded modes and ancillary modes.  Although used as a conceptual tool in the proof of Theorem~4, coherent photon injection has independent merit owing to its enabling preparation of high-photon-number Fock states from single-photon Fock states.  Thus we devote the next section to its description.\\

\textit{Coherent Photon Injection.}--- 
The coherent photon injection used in our universality proof is a generalization of a result from  Hubel \emph{et al}.~\cite{Jennewein2010}.  To illustrate how it works, suppose we start with the qubit-basis state $\ket{\tilde{1}} = \ket{0,0,1}$ from (\ref{qubitBasisEq}) with the goal of generating the qutrit-basis state $\ket{\tilde{2}} = \ket{0,0,2}$ from (\ref{Qutritbasis1}). Coherent photon injection accomplishes this task as follows.  We adjoin the $\ket{0,0,1}$ system with an ancillary pump mode (photon creation operator $\hat{a}_{p^\prime}^\dagger$) that has the same frequency as, but is orthogonally polarized to, the pump mode of $\ket{0,0,1}$.    We then turn on the $\chi^{(2)}$ interaction $\hat{G}_{2a}=\left[\hat{a}_s^{\dagger}\hat{a}_i^{\dagger}\hat{a}_{p^\prime} + \hat{a}_s\hat{a}_i\hat{a}_{p^\prime}^{\dagger}\right]$ between the original signal-idler modes and the ancillary pump mode 
to realize the transformation $ e^{i\pi\hat{G}_{2a}/2}  \ket{0,0,1}\ket{1}_a= \ket{1,1,1}\ket{0}_a $.  This coherent photon injection has transformed the qubit-basis state $\ket{\tilde{1}} = \ket{0,0,1}$ in the one-pump-photon subspace to the qutrit-basis state $\ket{\tilde{1}} = \ket{1,1,1}$ in the two-pump-photon subspace. A qutrit-basis $\chi^{(2)}$ gate can now rotate $\ket{\tilde{1}}$ to $\ket{\tilde{2}}$ in the two-pump-photon subspace~\cite{supp}.  Insofar as the pump mode is concerned, this overall procedure has converted a single-photon Fock-state input to a two-photon Fock-state output.  The injection process can now be repeated to transform $\ket{0,0,2}$ to  $\ket{1,1,2}$, after which a $\chi^{(2)}$-enabled rotation in the three-pump-photon subspace will yield $\ket{0,0,3}$. In this manner, high-photon-number Fock states can be prepared using only single-photon sources and full-quantum $\chi^{(2)}$ interactions.

\textit{Conclusions.}---
We have shown that universal optics-based quantum computation using only linear optics and $\chi^{(2)}$ interactions is possible in any $(n+1)$-dimensional qudit basis of the $n$-pump-photon subspace, with the natural basis being the three-mode Fock states $\{\ket{0,0,n},\ket{1,1,n-1},\ldots, \ket{0,0,n}\}$ of frequency-$\omega$, orthogonally-polarized signal and idler modes, and a frequency-$2\omega$ pump mode, all of which share a common spatial mode.  Our work extends the usual gate-model universality to the universality of $\chi^{(2)}$ Hamiltonian interactions in their irreducible subspaces. Such extension facilitates error correction for photon loss by providing a symmetry-operator formalism for hardware-efficient quantum error correction~\cite{ToBepublished}. Moreover, Lie algebraic understanding of $\chi^{(2)}$ interactions opens a path for defining an Abelian group that would enable fault-tolerant quantum computation that is robust to photon loss \emph{and} physical rotation errors.   To reach the end of that path, however, will require technology development.   

The resources required for our qudit-basis $\chi^{(2)}$ quantum computation are:  single photon sources, and linear optics, plus  $\chi^{(2)}$ interactions.  High-quality linear optics (dichroic mirrors and phase shifters) are already available, and high-efficiency quantum-state frequency conversion (the pumped $\chi^{(2)}$ interaction we need) have been demonstrated.  But, because currently available or demonstrated single-photon sources and full-quantum $\chi^{(2)}$ (SHG, SFG$_\pi$, and SPDC) interactions fall short of what our architectures require, continued advances in these technologies must occur before our quantum computation proposals become practical.  There is some reason for optimism in this regard, e.g., the efficiency of the $\chi^{(2)}$ nonlinearity has been improved from $ 10^{-7} $~\cite{Ramelow2011} to $10^{-1}$~\cite{Ramelow2017} in less than a decade.  Furthermore, state-of-the-art experimental realizations of strong $\chi^{(2)}$ interactions---including in solid-state circuits ~\cite{Long2008}, flux-driven Josephson parametric amplifiers~\cite{Devoret2010,Bergeal2010,Oliver2016}, superconducting resonator arrays \cite{Tian2011,Tian2012},  nondepleted four-wave-mixing-induced three-wave mixing in photonic microstructured fibers~\cite{Ramelow2011,Jennewein2014,meyer2015power}, $\chi^{(2)}$ interactions inside ring resonators~\cite{Sipe2007}, and nonlinear interactions in frequency-degenerate double-lambda systems~\cite{TwoLambda}---are closing the gap between theory and practical applications of full-quantum $\chi^{(2)}$ interactions. 

M.~Y.~N. and J.~H.~S. acknowledge support from Air Force Office of Scientific Research Grant No.~FA9550-14-1-0052.  M. Y. N. acknowledges  support from the Claude E. Shannon Research Assistantship.     I.~L.~C. acknowledges support from the National Science Foundation Center for Ultracold Atoms. M.~Y.~N. acknowledges early discussion with B. C. Sanders on Lie-group analysis of $ \chi^{(2)} $ interactions.

\begin{widetext}
\begin{center}
{\Large\bf Supplemental Material}
\end{center}

Here we present more information about the generalized sum-frequency generation (SFG$_\theta$) transformation, the proof of the main paper's Theorem~4, and the details of the $\ket{1,1,1}\rightarrow\ket{0,0,2}$ transformation needed to complete the coherent photon injection process for generating a two-photon Fock state from two single-photon Fock states. 

\section{Generalized Sum-Frequency Generation}
Consider the closed $t\ge 0$ joint-state evolution of the signal, idler, and pump modes in the main paper's one-pump-photon subspace, ${\rm Span}\{\ket{1,1,0},\ket{0,0,1}\}$, under the action of the $\chi^{(2)}$ Hamiltonian $i\hbar\kappa(\hat{a}_s^\dagger\hat{a}_i^\dagger\hat{a}_p-\hat{a}_s\hat{a}_i\hat{a}_p^\dagger)$, where $\kappa$ is real valued.  From~\cite{Ramelow2011,Niu2016UPDC} we have that this joint state satisfies
\begin{equation}
\ket{\psi(t)} = v_0(t)\ket{1,1,0} + v_1(t)\ket{0,0,1},\mbox{ for $t\ge 0$,}
\end{equation}
where 
\begin{equation}
\dot{v}_0(t)  = -\kappa v_1(t),\,\, \dot{v}_1(t) = \kappa v_0(t), \mbox{ for $t \ge 0$.}
\end{equation}
When $\ket{\psi(0)} = \ket{1,1,0}$, we find that
\begin{equation}
\ket{\psi(t)} = \cos(\kappa t)\ket{1,1,0} + \sin(\kappa t)\ket{0,0,1}, \mbox{ for $t \ge 0$}.
\end{equation}
Setting $\theta = \kappa t$ then gives us the generalized sum-frequency generation (SFG$_\theta$) transformation.

\section{Proof of the Main Paper's Theorem 4}\label{proof}
To complete the induction proof of the main paper's Theorem~4, we must show that $\chi^{(2)}$ interactions and linear optics are universal in the $(n+1)$-pump-photon subspace $\mathcal{H}_{n+1}$, given that these resources are universal in all $j$-pump-photon subspaces, $\mathcal{H}_j$, for $ 1\le j\leq n $.  Our proof has three steps, which make use of $\mathcal{H}_{n+1}$'s decomposition into its bulk states, $\mathcal{H}'_{n+1}\equiv {\rm Span}\{\ket{1,1,n},\ket{2,2,n-1},\ldots,\ket{n,n,1}\}$, and its boundary states, $\mathcal{H}^{\prime \perp}_{n+1} \equiv {\rm Span}\{\ket{0,0,n+1},\ket{n+1,n+1,0}\}$.  First, under the proof's premise, we prove the universality of $\chi^{(2)}$ interactions and linear optics in the bulk-state subspace $\mathcal{H}^\prime_{n+1}$. This step relies on the assumption that $\chi^{(2)}$ interactions and linear optics permit coherent photon subtraction/injection to be performed. Second, we show  how to achieve universality that spans $\mathcal{H}_{n+1}$ by including the boundary states, $\mathcal{H}^{\prime\perp}_{n+1}$, for that subspace.  Finally, we show that $\chi^{(2)}$ interactions suffice to realize the coherent photon  subtraction/injection operations used in Step~1.\\   

\noindent\emph{Step~1}: We begin by introducing additional  $\chi^{(2)}$ Hamiltonians between signal and idler modes with an ancillary  pump mode,
\begin{align}\label{G1p}
&\hat{G}_{1,p'}= \frac{i}{2}\left[\hat{a}_{s}^\dagger\hat{a}_{i}^\dagger\hat{a}_{p'} - \hat{a}_{s}\hat{a}_{i}\hat{a}_{p'}^\dagger\right]  ,\\\label{G2p}
&\hat{G}_{2,p'}= \frac{1}{2}\left[\hat{a}_{s}^\dagger\hat{a}_{i}^\dagger\hat{a}_{p'} + \hat{a}_{s}\hat{a}_{i}\hat{a}_{p'}^\dagger \right],
\end{align}
as well as between ancillary signal and idler modes and the pump mode,
\begin{align}\label{G1si}
&\hat{G}_{1,s'i'}= \frac{i}{2}\left[\hat{a}_{s'}^\dagger\hat{a}_{i'}^\dagger\hat{a}_{p} - \hat{a}_{s'}\hat{a}_{i'}\hat{a}_{p}^\dagger\right]  ,\\\label{G2si}
&\hat{G}_{2,s'i'}= \frac{1}{2}\left[\hat{a}_{s'}^\dagger\hat{a}_{i'}^\dagger\hat{a}_{p} + \hat{a}_{s'}\hat{a}_{i'}\hat{a}_{p}^\dagger \right].
\end{align}
Here we have taken $\kappa = 1$, and we have assumed that: (1) the ancillary pump mode and the pump mode have the same $2\omega$ frequency, but are orthogonally polarized; and (2) the ancillary signal and idler modes have frequencies $\omega_{s'} = \omega + \Delta\omega$ and $\omega_{i'} = \omega - \Delta\omega$, so that they are orthogonal to each other and to the signal and idler modes, which both have frequency $\omega$.

Next, we assume that linear optics plus the $\chi^{(2)}$ interactions $\hat{G}_{1,p'}$ and $\hat{G}_{2,p'}$ allow us to realize the unitary transformation $\hat{U}_{sip'}$ that accomplishes the following coherent photon subtraction on $\mathcal{H}_{n+1}$, 
\begin{equation}\label{Upprime}
\hat{U}_{sip'}\ket{j,j,n+1-j}\ket{0,0}_{s'i'} \ket{n-3}_{p'}=\ket{j-1,j-1,n+1-j}\ket{0,0}_{s'i'} \ket{n-2}_{p'},\mbox{ for $1\le j \le n, n \ge 3$.}
\end{equation}
Likewise, we assume that linear optics plus the $\chi^{(2)}$ interactions $\hat{G}_{1,s'i'}$ and $\hat{G}_{2,s'i'}$ allow us to realize the unitary transformation $\hat{U}_{ps'i'}$ that accomplishes the following coherent photon subtraction on $\mathcal{H}_{n+1}$,
\begin{equation}\label{Usiprime}
\hat{U}_{ps'i'} \ket{j-1,j-1,n+1-j}\ket{0,0}_{s'i'} \ket{n-2}_{p'}=\ket{j-1,j-1,n-j}\ket{1,1}_{s'i'} \ket{n-2}_{p'}, \mbox{ for $1\le j \le n, n \ge 2$.}
\end{equation}
Concatenating these transformations then converts any state in $\mathcal{H}'_{n+1}$ into a corresponding state in $\mathcal{H}_{n-1}$ by means of the basis transformation
\begin{equation}
\hat{U}_{ps'i'}\hat{U}_{sip'}\ket{j,j,n+1-j}\ket{0,0}_{s'i'} \ket{n-3}_{p'} = \ket{j-1,j-1,n-j}\ket{1,1}_{s'i'} \ket{n-2}_{p'}, \mbox{ for $1\le j \le n, n \ge 3$.}
\end{equation}
This concatenation, however, will also unavoidably impact the $\mathcal{H}_{n+1}$ boundary basis states $\ket{0,0,n+1}$ and $\ket{n+1,n+1,0}$, i.e., there will be $\{c_k\}$ and $\{d_k\}$ such that
\begin{align}\label{effect1}
&\hat{U}_{ps'i'}\hat{U}_{sip'}\ket{0,0,n+1}\ket{0,0}_{s'i'} \ket{n-3}_{p'}=\sum_{k=0}^{n+1} c_k\ket{0,0,n+1-k}\ket{k,k}_{s'i'} \ket{n-3}_{p'},\\\label{effect2}
&\hat{U}_{ps'i'}\hat{U}_{sip'}\ket{n+1,n+1,0}\ket{0,0}_{s'i'} \ket{n-3}_{p'}=\sum_{k=0}^{n+1} d_k\ket{n+1-k, n+1-k, 0}\ket{0,0}_{s'i'} \ket{n-3+k}_{p'}.
\end{align}

Note that the transformations in Eqs.~(\ref{effect1}) and (\ref{effect2}) do not contain any states with a $\ket{1,1}_{s'i'} \ket{n-2}_{p'}$ component, which is in the $(n-1)$-pump-photon subspace of the ancillary signal, idler, and pump modes.  We can thus realize any gate $\hat{U}^{\rm target}_n$ that acts on the bulk-state subspace $\mathcal{H}^\prime_{n+1}$, without affecting the boundary basis states, using only linear optics and $\chi^{(2)}$ interactions.  In particular, let $\hat{\tilde{U}}^{\rm target}_n$ be the mapping of $\hat{U}^{\rm target}_n$ to a unitary that acts on $\mathcal{H}_{n-1}$.  Then, from Theorem~4's premise, we know we can realize the controlled unitary gate $\Lambda[\hat{\tilde{U}}^{\rm target}_n]$ that is conditioned on the ancillary modes' being in their $\ket{1,1}_{s'i'} \ket{n-2}_{p'}$ state:
\begin{equation}
\Lambda[\hat{\tilde{U}}^{\rm target}_n]  =  \hat{\tilde{U}}^{\rm target}_n\otimes \ket{1,1}_{s'i'} \ket{n-2}_{p'\,p'}\!\bra{n-2}_{\,s'i'}\!\bra{1,1}  + \hat{I}_{n-1}\otimes\sum_{\{j,l\}\neq \{1,n-2\}}\ket{j,j}_{s'i'} \ket{l}_{p'\,p'}\bra{l}_{\,s'i'}\bra{j,j},
\end{equation}
where $\hat{I}_{n-1}$ is the $\mathcal{H}_{n-1}$ identity operator.  We then undo the effects in Eqs.~(\ref{effect1}) and (\ref{effect2}) by conjugating with 
$\hat{U}_{ps'i'}\hat{U}_{sip'}$, and so obtain
\begin{equation}
\hat{U}^{\rm target}_n =  (\hat{U}_{ps'i'}\hat{U}_{sip'})^\dagger  \Lambda[\hat{\tilde{U}}^{\rm target}_n]\hat{U}_{ps'i'}\hat{U}_{sip'}.
\end{equation}
This result completes Step~1, because $\chi^{(2)}$ interactions and linear optics enable realization of the coherent photon injection operations embodied by $(\hat{U}_{ps'i'}\hat{U}_{sip'})^\dagger$ as they are the inverses of the coherent photon subtraction operations in $\hat{U}_{ps'i'}\hat{U}_{sip'}$, whose realizations with those resources we have already assumed (and will prove in Step~3).\\

\noindent\emph{Step~2}:  
 We have shown that universal gates in $\mathcal{H}_{n+1}'$ can be implemented from $\chi^{(2)}$ interactions and linear optics, given that such resources suffice to realize universal gates in $\mathcal{H}_j$ for $1\le j\leq n$ and to do coherent photon subtraction and injection.  It thus remains for us to show that this universality result can be extended to $\mathcal{H}_{n+1}$ under that same premise, i.e., we must now include the boundary states, $\mathcal{H}^{\prime\perp}_{n+1}$, that are in $\mathcal{H}_{n+1}$ but not in $\mathcal{H}_{n+1}'$.  To do so we start by decomposing the restrictions to $\mathcal{H}_{n+1}$ of the main paper's $\chi^{(2)}$ Hamiltonians $\hat{G}_1$ and $\hat{G}_2$ into their components that act within $\mathcal{H}^\prime_{n+1}$ and those that act between the boundary states, $\mathcal{H}^{\prime\perp}_{n+1}$, and the bulk states, $\mathcal{H}^\prime_{n+1}$.  Using $\hat{\tilde{G}}_1$ and $\hat{\tilde{G}}_2$ to denote the $\mathcal{H}_{n+1}$-restricted Hamiltonians, and assuming $\kappa=1$, we can show that
\begin{align}
\hat{\tilde{G}}_1\label{PAULIXYG1}
&=\frac{1}{2}\left[ \sqrt{n+1}\,\hat{\sigma}_{n+1,n}^y + (n+1) \hat{\sigma}_{1,0}^y +\sum_{k=1}^{n-1}(n+1-k)\sqrt{k+1} \,\hat{\sigma}_{k+1,k}^y \right], \\
\hat{\tilde{G}}_2\label{PAULIXYG2}
&=\frac{1}{2}\left[ \sqrt{n+1}\,\hat{\sigma}_{n+1,n}^x + (n+1)\hat{\sigma}_{1,0}^x +\sum_{k=1}^{n-1}(n+1-k)\sqrt{k+1} \,\hat{\sigma}_{k+1,k}^x \right], 
\end{align}
where $\hat{\sigma}_{k+1,k}^x$ and $\hat{\sigma}_{k+1,k}^y$ are the Pauli $\hat{X}$ and $\hat{Y}$ operators between the basis states $\ket{n-k,n-k,k+1}$ and $\ket{n+1-k,n+1-k,k}$.  

Next, from Step~1, we know that we can implement the $\mathcal{H}^\prime_{n+1}$ unitary gates $\hat{U}_1(\theta)=\exp\!\left[-i\theta\left(\sum_{k=1}^{n-1}(n+1-k)\sqrt{k+1}\, \hat{\sigma}_{k+1,k}^y  \right)   \right]$ and $\hat{U}_2(\theta)=\exp\!\left[-i\theta\left( \sum_{k=1}^{n-1}(n+1-k)\sqrt{k+1}\, \hat{\sigma}_{k+1,k}^x  \right)   \right]$ with only the presumed resources. Then, in order to implement $\hat{\tilde{G}}_1$ and $\hat{\tilde{G}}_2$ using only two-dimensional subspaces, we leverage the Trotter formula to obtain the following entangling operators between $\mathcal{H}^\prime_{n+1}$ and the two boundary basis states $\ket{n-k,n-k,k+1}$ and $ \ket{n+1-k,n+1-k,k}$:
\begin{align}\nonumber
 \hat{V}_1(\theta) &= \exp\!\left[i\theta  ( \sqrt{n+1}\,\hat{\sigma}_{n+1,n}^y + ( n+1) \hat{\sigma}_{1,0}^y)\right]\\
&=\exp\!\left[ 2i\theta \hat{G}_1 -i\theta\left(  \sum_{k=1}^{n-1}(n+1-k)\sqrt{k+1}\, \hat{\sigma}_{ k+1,k}^y  \right) \right]=\lim_{m\to \infty}\left[  e^{ 2i\theta \hat{G}_1/m} \hat{U}_1(\theta)^{1/m}\right]^m,\\\nonumber
 \hat{V}_2(\theta)&=\exp\!\left[i\theta  ( \sqrt{n+1}\,\hat{\sigma}_{n+1,n}^x + ( n+1) \hat{\sigma}_{1,0}^x)\right]\\
 &=\exp\!\left[2i\theta \hat{G}_2 -i\theta\left(  \sum_{k=1}^{n-1}(n+1-k)\sqrt{k+1}\,\hat{\sigma}_{ k+1,k}^x  \right)\right]=\lim_{m\to \infty}\left[  e^{ 2i\theta \hat{G}_2/m } \hat{U}_2(\theta)^{1/m}\right]^m.
\end{align}
At this point, we can choose rotation angles such that $\hat{V}_1(\theta_1)\hat{V}_1(\theta_2)$ and  $\hat{V}_2(\theta_3)\hat{V}_2(\theta_4)$ respectively construct individual Pauli $\hat{X}$ and $\hat{Y}$ rotations either between $ \ket{n+1,n+1,0}$ and $\ket{n ,n ,1}$,  or   between $ \ket{0,0,n+1}$ and  $\ket{1,1,n}$.  Six of these rotations, with different angles, can then be used to realize any SU(2) rotation between $\ket{0,0,n+1}$ and $\ket{1,1,n}$, or  between $\ket{n+1,n+1,0} $ and $\ket{n ,n ,1}$~\cite{Child1998,MikeIke,Kemp2001,Freedman2002}.  So, because any unitary transformation can be decomposed into products of unitary transformations between two neighboring basis states of any chosen order~\cite{MikeIke},  we have proven that $\chi^{(2)}$ interactions and linear optics are sufficient to realize universal gates in $\mathcal{H}_{n+1}$ \emph{assuming} that these resources suffice to realize such gates in $\mathcal{H}_j$ for $1\le j \le n$, \emph{and} that they also suffice for realizing the coherent photon subtraction/injection operations that were employed in Step~1.\\

\noindent\emph{Step~3}:  With Step~2 in hand, completing Theorem~4's induction proof only requires showing that $\chi^{(2)}$ interactions can be used to implement the coherent photon subtraction operations---$\hat{U}_{sip'}$ and $\hat{U}_{s'i'p}$ from Eqs.~(\ref{Upprime}) and (\ref{Usiprime})---that were employed in Step~1. (Step~1 also used the coherent photon injection operations $\hat{U}_{sip'}^\dagger$ and $\hat{U}_{s'i'p}^\dagger$, but their realizations are merely inverses of the unitaries for their associated subtraction processes.)  In particular, we need only demonstrate that the available resources are universal in the joint Hilbert space $\mathcal{H}^{\rm joint}_n \equiv \left(\oplus_{j=1}^n\mathcal{H}_j'\right)  \otimes \mathbb{C}^2_{s'i'}  \otimes \mathbb{C}^2_{p'}= \text{Span}\{\ket{j,j, k}\ket{l,l}_{s'i'}\ket{h}_{p'} : 1 \leq j + h \leq n, \,1 \leq  k + l\leq n, \, \,\, h, l \in\{0,1\} \} $. 

Let $\mathcal{U}_{\rm sum} $ be the unitary group in $\oplus_{j=1}^n\mathcal{H}_j^\prime$, and let $\mathcal{S}$ be the unitary group generated by linear optics and $\chi^{(2)}$ interactions. By the induction proof's premise we know that  $\mathcal{U}_{\rm sum} \subset \mathcal{S}$. Moreover, because linear optics alone is universal in $\mathbb{C}^2_{s'i'}$, and so too is it universal in $\mathbb{C}^2_{p'}$, we have that $\mathcal{U}_{\rm sum}\otimes \mathcal{U}_{s'i'}\otimes \mathcal{U}_{p'} \subset \mathcal{S}$, where $\mathcal{U}_{s'i'}$ is the unitary group in $\mathbb{C}^2_{s'i'}$ and $\mathcal{U}_{p'}$ is the unitary group in $\mathbb{C}^2_{p'}$. The proof from Ref.~\cite{Harrow2009} will now establish the universality of $\mathcal{S}$ in $\mathcal{H}^{\rm joint}_{n}$ if we can show there are imprimitive gates $\hat{V}_{s'i'}$ and $\hat{V}_{p'}$ in $\mathcal{S}$  that perform entangling operations between  subspaces  $\oplus_{j=1}^n\mathcal{H}_j^\prime $ and $\mathbb{C}^2_{s'i'}$ and between $\oplus_{j=1}^n\mathcal{H}_j^\prime $ and $\mathbb{C}^2_{p'}$, viz., for $\ket{\psi}_n\in \oplus_{j=1}^n\mathcal{H}_j^\prime$, $\ket{\psi'}_{s'i'} \in  \mathbb{C}^2_{s'i'}$, and $\ket{\psi'}_{p'} \in  \mathbb{C}^2_{p'}$, we have that $\hat{V}_{s'i'}\ket{\psi}_n\ket{\psi'}_{s'i'}$ and $\hat{V}_{p'}\ket{\psi}_n\ket{\psi'}_{p'}$ are entangled states.  Below we will establish the existence of $\hat{V}_{p'}$; a similar argument, which we omit, will do the same for $\hat{V}_{s'i'}$. 

Consider the initial product state, 
\begin{align}
\ket{\psi}_n\ket{\psi'}_{p'} =\left[\sum_{k=1}^{n-1}\sum_{j=0}^{k-1}\alpha_{j}\ket{j,j,k-j} \right] \left[ \sum_{q=0}^1\beta_{q} \ket{q}_{p'}   \right],
\end{align}
in $(\oplus_{j=1}^n\mathcal{H}_j^\prime)\otimes \mathbb{C}^2_{p'}$.
Based on the universality of  $\chi^{(2)}$ interactions plus linear optics in $j$-pump-photon subspaces with $1\le j\le n$, there exists a unitary $\hat{U}'_{sip'}$,  generated by $\hat{G}_{1,p'}$ and $\hat{G}_{2,p'}$, that transforms $\ket{\psi}_n\ket{\psi'}_{p'}$  into 
\begin{align}
\hat{U}'_{sip'}\ket{\psi}_n\ket{\psi'}_{p'}=& \sum_{k=1}^{n-1}\sum_{j=0}^{k-1}(\alpha_{j}\beta_{1}\gamma_{j,1} +\alpha_{j+1}\beta_{0}\gamma_{j+1,0}) \ket{j+1,j+1,k-j}\ket{0}_{p'} \nonumber \\
& + \sum_{k=1}^{n-1}\sum_{j=1}^{k-1}(\alpha_{j}\beta_{0}\gamma_{j,0} +\alpha_{j-1}\beta_{1}\gamma_{j-1,1}) \ket{j-1,j-1,k-j}\ket{1}_{p'} \label{Uprime}
\end{align}
where the $\{\gamma_{j,q}\}$ are determined by the $\chi^{(2)}$  Hamiltonian evolution between signal, idler and ancillary pump modes.   Equation~(\ref{Uprime}) generates entanglement whenever $\gamma_{j,q}\neq \gamma_{j',q'}$ for any $ j\neq j' $ or $ q\neq q' $.  Such will always be the case with an appropriate rotation angle $\theta$ of the $\chi^{(2)}$ Hamiltonian evolution  generated by  Eqs.~(\ref{G1p}) and (\ref{G2p}). Thus $\chi^{(2)}$ interactions and linear optics are universal in $\mathcal{H}^{\rm joint}_n$ and hence sufficient to implement coherent photon subtraction/injection,  completing the proof of the main paper's Theorem~4.

\section{$\chi^{(2)}$-enabled approach for realizing the $\ket{1,1,1}\to \ket{0,0,2}$ transformation}\label{convert}

The main paper's description of using coherent photon injection to generate a two-photon Fock state from two single-photon Fock states via coherent photon injection required a $\chi^{(2)}$-enabled approach to accomplishing the $\ket{1,1,1}\to \ket{0,0,2}$ transformation.  Here we will provide a suitable implementation for that transformation. The first step is to apply the $\chi^{(2)}$ Hamiltonian $\hat{G} = i\hbar\kappa(\hat{a}_s^\dagger\hat{a}_i^\dagger\hat{a}_p-\hat{a}_s\hat{a}_i\hat{a}_p^\dagger)$ to the $\ket{1,1,1}$ state for time $t= 2\pi/3\kappa\sqrt{6}$. Using the $\mathcal{H}_2$ evolution equation from~\cite{Niu2016UPDC}, we find that
\begin{align}
\ket{\psi_1} \equiv e^{-i2\pi\hat{G}/3\kappa\sqrt{6}}\ket{1,1,1}= -\frac{1}{2}\ket{0,0,2} - \frac{1}{2}\ket{1,1,1} + \frac{1}{\sqrt{2}}\ket{2,2,0}.
\end{align} 
The second step is to send the preceding quantum state through the optical circuit shown in Fig.~\ref{fig1}.  Here, the pump mode is routed to the upper rail by the first dichroic mirror (DM), while the second-harmonic generation (SHG) block converts a two-photon Fock-state idler (if present) to a single-photon Fock-state at the pump frequency.  That pump-frequency single-photon state is converted back into a two-photon Fock-state idler by spontaneous parametric downconversion (SPDC).  In that process it accumulates a $\pi$-rad Berry's phase.  The circuit is completed by using a DM to put the pump mode on the output rail.  The net effect of this circuit is then to transform $\ket{\psi_1}$ to 
\begin{align}
\ket{\psi_2}=-\frac{1}{2}\ket{0,0,2} - \frac{1}{2}\ket{1,1,1} - \frac{1}{\sqrt{2}}\ket{2,2,0}.
\end{align}
\begin{figure}[H]
\begin{center}
\includegraphics[width=0.4\linewidth]{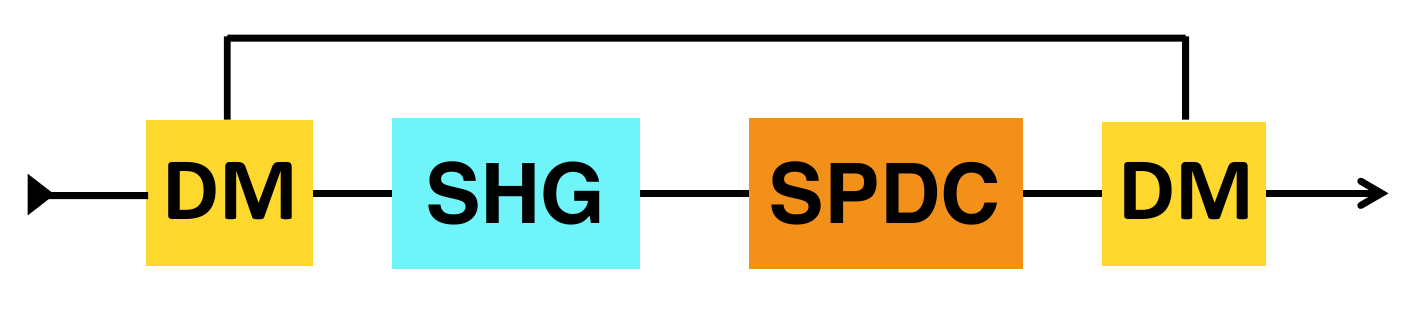}
\caption{Optical circuit for the second step of the $\ket{1,1,1}\to \ket{0,0,2}$ transformation.  DM:  dichroic mirror.  SHG:  second-harmonic generation. SPDC: spontaneous parametric downconversion.} \label{fig1}
\end{center}
\end{figure}
The last step is to evolve $\ket{\psi_2}$ under $\hat{G}$ for time $t=2\pi/3\kappa\sqrt{6}$ to obtain
\begin{align}
\ket{\psi_3} \equiv e^{-i2\pi \hat{G}/3\kappa \sqrt{6}}\ket{\psi_2}=-\ket{0,0,2}, 
\end{align}
which, because the global phase is irrelevant, completes the desired transformation. 

\end{widetext}


\begin{thebibliography}{44}
\bibitem{LieReference}
Readers unfamiliar with Lie algebras and Lie groups should consult, e.g.,~\cite{LieText} for more information about them.
\end{thebibliography}
\end{document}